\pdfoutput=1
\documentclass[twocolumn,prb,showpacs,aps,superscriptaddress]{revtex4-1}

\usepackage[english]{babel}
\usepackage[ansinew]{inputenc}
\usepackage{times}
\usepackage{graphicx}
\usepackage{graphics}
\usepackage{amsmath}
\usepackage{amsfonts}
\usepackage{amssymb}
\usepackage{amsbsy}
\usepackage{epstopdf}
\usepackage{makeidx}
\usepackage{color}
\usepackage{pgf}
\usepackage{bm}
\usepackage{tikz} 
\usepackage[normalem]{ulem}
\usepackage{multirow,bigstrut}

\newcommand{\be}{\begin{equation}}
\newcommand{\ee}{\end{equation}}
\newcommand{\bea}{\begin{eqnarray}}
\newcommand{\eea}{\end{eqnarray}}

\makeindex
\begin{document}

\title{Data-driven Material Models for Atomistic Simulation}

\author{M. A. Wood}
\affiliation{Center for Computing Research, Sandia National Laboratories, Albuquerque, New Mexico 87185, USA}
\author{M. A. Cusentino}
\affiliation{Center for Computing Research, Sandia National Laboratories, Albuquerque, New Mexico 87185, USA}
\author{B. D. Wirth}
\affiliation{University of Tennessee, Knoxville, Tennessee 37996, USA}
\author{A. P. Thompson}
\affiliation{Center for Computing Research, Sandia National Laboratories, Albuquerque, New Mexico 87185, USA}

\date{\today}

\begin{abstract}
The central approximation made in classical molecular dynamics simulation of materials is the interatomic potential used to calculate the forces on the atoms.  
Great effort and ingenuity is required to construct viable functional forms and find accurate parameterizations for potentials using traditional approaches.
Machine-learning has emerged as an effective alternative approach to develop accurate and robust interatomic potentials.  
Starting with a very general model form, the potential is learned directly from a database of electronic structure calculations and therefore can be viewed as a multiscale link between quantum and classical atomistic simulations.
Risk of inaccurate extrapolation exists outside the narrow range of time- and length-scales where the two methods can be directly compared.
In this work, we use the Spectral Neighbor Analysis Potential (SNAP) and show how a fit can be produced with minimal interpolation errors which is also robust in extrapolating beyond training. 
To demonstrate the method, we have developed a new tungsten-beryllium potential suitable for the full range of binary compositions. 
Subsequently, large-scale molecular dynamics simulations were performed of high energy Be atom implantation onto the (001) surface of solid tungsten. 
The new machine learned W-Be potential generates a population of implantation structures consistent with quantum calculations of defect formation energies.  
A very shallow ($< 2$nm) average Be implantation depth is predicted which may explain ITER diverter degradation in the presence of beryllium.
\end{abstract}


\maketitle
\section{\label{intro}Introduction}
Over the past few decades the rapid advancements and availability of computing technologies has changed the way research is conducted in many areas of science and engineering. 
Modern supercomputing systems enable researchers to perform hundreds or thousands of virtual experiments before setting foot in a traditional laboratory.
One of the main advances with these computational efforts has been the curation of results into extensive open source databases, enabling the data to be used to drive materials discovery and model development, often in ways never intended by the originators.\cite{saal2013materials,jain2013commentary,zhuang2013computational,villars2006asm,kim2018polymer}
A recent trend in material science is the adoption of data science techniques to derive new understanding of material properties from modeling and simulation.
In the present work we use machine learning to bridge between quantum and classical atomistic simulation methods, which can be viewed as a particular
case of data-driven materials modeling.

For materials behavior that originates at the atomic scale, molecular dynamics (MD) is a powerful and popular computational tool.
This work highlights a computational multiscale approach where a database of electronic structure calculations is translated into a classical interatomic potential(IAP) for MD. Calculation of forces using an IAP is many orders of magnitude more computationally efficient than using quantum electronic structure methods such as density funtcional theory (DFT), while capturing the same essential physics.
It is important to realize that the key approximation made in MD simulations is the interatomic potential.
For this reason, great care needs to be taken in the IAP construction, as well as in interpretation of simulation results that are computed from a given potential.
There are many different mathematical forms that can be used to construct an interatomic potential, many of these use physics and chemistry as a model\cite{daw1993embedded,baskes1992modified,sinnott2012three,liang2013reactive,tersoff1988empirical,lennard1931cohesion} to determine the forces on the atoms. 
However, there is a recent trend of relying on machine-learning approaches\cite{botu2016machine,bartok2013representing,rupp2018guest,schutt2018schnet} to construct an IAP that can significantly decrease the time investment needed while simultaneously improving the accuracy with respect to electronic structure predictions \cite{chen2017accurate,bartok2018machine,wood2018extending}.
Additionally, this data-science approach to an IAP can be applied to materials with complex bonding characteristics which are challenging for traditional potentials\cite{plimpton2012computational,deng2019electrostatic,li2018quantum}.  

An example case where traditional IAP have trouble representing atomic interactions is the W-Be material system which is of relevance to modeling plasma material interactions for fusion devices.  
For the ITER reactor, beryllium and tungsten have been chosen as the first wall and diverter materials, respectively.  They have already been used in experimental fusion reactors\cite{federici2003key,brezinsek2015plasma}.  
Due to the low atomic number of beryllium and its favorable thermal conductivity, it is a suitable material for the first wall where impurity transport into the plasma is a concern\cite{federici2001plasma}.  
On the other hand, the divertor region receives the highest ion and heat fluxes, on the order of 10$^{24}$~m$^{-2}$s$^{-1}$ and 10~MWm$^{-2}$ respectively\cite{federici2001plasma}.
Tungsten has been chosen for these extreme conditions, because of its high melting point, good thermal conductivity, and low sputtering yield\cite{federici2001plasma}.
While the divertor region is expected to receive the highest ion and heat fluxes, some beryllium will be eroded from the first wall and deposited onto the divertor material\cite{brezinsek2013fuel,brezinsek2013residual}.  
This deposition of beryllium into the tungsten surface could lead to the formation of stable W-Be intermetallic compounds with much lower melting points than pure tungsten\cite{okamoto1991phase}.  Any reduction in the melting point of the divertor material could lead to a drastic increase in sputtering yield and deterioration of the divertor performance.
For this reason it is important to understand in detail how beryllium implants into tungsten and what types of mixed phases are formed near the surface.

Multiple experiments at PISCES-B \cite{doerner2005beryllium,doerner2007implications,baldwin2007w} of beryllium seeded deuterium plasma exposure of tungsten have been conducted to assess mixed material effects on deuterium retention and intermetallic formation.  
For plasmas containing as little as 0.1\% beryllium SEM images of the tungsten targets show both layers and deposits of various W-Be intermetallics including WBe$_{12}$\cite{baldwin2007w}.  
Additional XPS measurements indicate the formation of WBe$_2$ during the annealing process from 300 K up to 970 K \cite{linsmeier2007binary}.  
These experiments indicate that W-Be intermetallic formation in the diverter of ITER can occur and correspondingly, additional experimental and modeling efforts are needed to understand the underlying physical processes and mechanisms leading to intermetallic formation.  

Molecular dynamics is well suited to modeling these effects.
However, there are not many IAP developed for tungsten and beryllium and their accuracy is limited for this particular application.  
While many potentials exist for tungsten\cite{cusentino2015comparison}, only one exists for modeling W and Be\cite{bjorkas2010w}, which is a Tersoff style bond order potential (BOP)\cite{tersoff1988empirical}.  
This potential has been used to study both beryllium implantation in tungsten \cite{lasa2014atomistic} and mixed beryllium-deuterium implantation in tungsten \cite{lasa2014effect}.  
However, this potential form is not robust enough to capture the complex interactions between tungsten, beryllium, and their intermetallic structures.
In this article we show how the Spectral Neighbor Analysis Potential (SNAP) machine learning technique can be used to derive an IAP for W-Be that is capable of studying in detail these mixed material interactions.


\section{\label{model} Potential Energy Model}

An interatomic potential should accurately represent the many-body potential energy surface as a function of the local environment around an atom.   
By only considering neighbors within a distance of approximately 1~nm, classical MD simulations using parallel algorithms can be scaled far beyond what is possible for electronic structure codes.  This remains true for the data-science inspired potentials.\cite{plimpton1995computational}
Machine learned interatomic potentials (ML-IAP) can be distinguished from one another based on three key factors; regression technique, choice of descriptors and energy model form. 
Many of the recently developed machine learned interatomic potentials can be placed on a continuous scale of being more physical- or data-science based.\cite{ramprasad2017machine} 
Deep neural networks (NN) with simple descriptors and activation functions\cite{behler2007generalized,behler2016perspective,lubbers2018hierarchical,wang2018deepmd} directly exploit the recent advances in the field of data-science.
The key advantage of NN-based potentials is the immense flexibility of the model to capture even the most subtle features of the training data.
A limiting factor of these ML-IAP is the uncertainty in extrapolating beyond the training data to predict energies and forces in previously unseen atomic environments.
Non-parametric regression methods like Gaussian process\cite{bartok2010gaussian} or kernel ridge regression\cite{huan2017universal} use physically motivated kernels like local atom densities or bond topology and are toward the center of this scale\cite{botu2017study}.
The Spectral Neighborhood Analysis Potential (SNAP)\cite{thompson2015spectral}, which is used in this work, is more strongly physics-based, due to its use of the bispectrum as descriptors, which are closely related to invariants of the radial and angular basis functions of the atomic cluster expansion that is the natural description of the bonding environment around an atom\cite{drautz2019, seko2019}.
Additionally, for simplicity and computational efficiency, SNAP uses linear regression in order to decouple the computational cost at MD runtime from the details of the training set used. 

\subsection{Spectral Neighborhood Analysis Potential}
\newcommand{\hcoeff}[9]{H\!\!{\tiny{\begin{array}{l}#1 #2 #3 \\ #4 #5 #6 \\ #7 #8 #9 \end{array}}}}

We outline here the structure of the SNAP ML-IAP in terms of the underlying descriptor space.\cite{thompson2015spectral}
The total potential energy of a configuration of atoms is first written as the sum of SNAP energy contributions associated with each atom, combined with a reference
potential
\begin{equation}
E({\bf r}^{N})=E_{ref}({\bf r}^{N})+\sum_{i=1}^{N}E_{\scriptscriptstyle{SNAP}}^i,
\label{snapE}
\end{equation}
where ${\bf r}^N$ is the vector of $N$ atom positions in the configuration. 
$E$ and $E_{ref}$ are the total and reference potential energies, respectively.
$E_{\scriptscriptstyle{SNAP}}^i$ is the SNAP potential energy associated with a particular atom $i$, and depends only on the relative positions of its neighbor atoms.
Including a reference potential is advantageous because it can correctly represent known limiting cases of atomic interactions, leaving the SNAP contribution to capture many-body effects.
The ZBL pair potential\cite{biersack1982stopping} is a convenient choice, because it captures the known short-range repulsive interactions between atomic cores that are not well represented by quantum calculations. 

The construction of the SNAP component of the potential energy in terms of the bispectrum components follows the same approach described in
Ref. [\onlinecite{thompson2015spectral}], which we briefly summarize here.
The SNAP formulation begins with a very general characterization of the neighborhood of an atom.
The density of neighbor atoms at location $\textbf{r}$ relative to a central atom $i$ located at the origin can be considered as a sum of $\delta$-functions located in a three-dimensional space:
\begin{equation}
\rho_i ({\bf r}) = \delta({\bf r}) + \sum_{r_{i'} < R_{ii'}}{f_c(r_{i'}) w_{i'} \delta({\bf r}-{\bf r}_{i'})}
\label{density}
\end{equation}
where ${\bf r}_{i'}$ is the position of neighbor atom $i'$ relative to central atom $i$.  
The $w_{i'}$ coefficients are dimensionless weights that are chosen to distinguish atoms of different types, while the central atom is arbitrarily assigned a unit weight.  
This sum is over all atoms $i'$ within the cutoff distance $R_{ii'}$ that is defined in terms of the effective radii of the two atoms
\begin{equation}
R_{ii'} = \alpha(R_i+R_{i'}),
\label{mixing}
\end{equation}
where $\alpha$ is a universal scale factor and $R_i$ and $R_{i'}$ are the effective radii of atom $i$ and $i'$ respectively. 
The switching function $f_c(r)$ ensures that the contribution of each neighbor atom goes smoothly to zero at $R_{ii'}$.  

Typically, this density function is expanded in an angular basis of spherical harmonics combined with an orthonormal radial basis. \cite{bartok2013representing}
Instead, we use an idea originally proposed by Bart{\'{o}}k et al.\cite{bartok2010gaussian}, in which the radial coordinate $r$ is mapped on to a third angular coordinate $\theta_0 = \theta_0^{max} r / R_{ii'}$.  
Each neighbor position $(r, \theta, \phi)$ is mapped to $(\theta_0, \phi, \theta)$, a point on the unit 3-sphere.  
The natural basis for functions on the 3-sphere is formed by the 4D hyperspherical harmonics $U^j_{m,m'}(\theta_0,\theta,\phi)$, defined for $j=0,\frac{1}{2},1,\ldots$ and  $m,m' = -j,-j\!+\!1,\ldots,j\!-\!1,j$~\cite{varshalovich1988quantum}.  
The neighbor density function can now be expanded in the basis of hyperspherical harmonics $U^j_{m,m'}$.  
Because the neighbor density is a weighted sum of $\delta$-functions, each expansion coefficient is a sum over discrete values of the corresponding basis function evaluated at each neighbor position 
\begin{equation}
u^j_{m,m'} = U^j_{m,m'}(0) + \!\!\!\!\!\!\!\!\sum_{\tiny{r_{i'} < R_{ii'}}}{\!\!\!\!f_c(r_{i'}) w_{i'} U^j_{m,m'}(\theta_0,\theta,\phi)} 
\label{eq:u}
\end{equation}
The bispectrum components are formed as the scalar triple products of the expansion coefficients
\begin{equation}
B_{j_1,j_2,j}  = \\
\!\!\!\!\sum_{m,m'} u^{j*}_{m,m'} \!\!\!\!\sum_{\tiny\begin{array}{l} \!\!m_1,m'_1 \\ m_2,m'_2 \end{array}}
\hcoeff{j}{m}{m'}{j_1}{\!m_1}{\!m'_1}{j_2}{m_2}{m'_2}
u^{j_1}_{m_1,m'_1} u^{j_2}_{m_2,m'_2}
\label{eq:bispectrum}
\end{equation}
where * indicates complex conjugation and the constants
$\hcoeff{j}{m}{m'}{j_1}{\!m_1}{\!m'_1}{j_2}{m_2}{m'_2}$
are Clebsch-Gordan coupling coefficients for the hyperspherical harmonics.
Importantly, the bispectrum components are real-valued and invariant under rotation~\cite{bartok2010gaussian}.  
They are also symmetric in the three indices $j_1, j_2, j$ up to a normalization factor.~\cite{thompson2015spectral}
They characterize the strength of density correlations at three points on the 3-sphere.
The lowest-order components describe the coarsest features of the density function, while higher-order components reflect finer detail. 
The number of distinct bispectrum components with indices $j_1, j_2, j$ less than or equal to $J$ increases as $J^3$.  
For a particular choice of $J$, we can list the $K$ bispectrum components in some arbitrary order as ${B}_{1},\ldots,{B}_{K}$.  
The SNAP energy of an atom is written as a linear function of the bispectrum components 
\begin{eqnarray}
E_{\scriptscriptstyle{SNAP}}^i & = & {\beta}_{0}+\sum_{k=1}^{K}{\beta_k}({B}_{k}^{i}-{B}_{k{0}}^{i}) \\
& = & {\beta}_{0}+\boldsymbol\beta\cdot{\bf B}^{i}
\label{eq2}
\end{eqnarray}
where ${B}_{k}^{i}$ is the $k$th bispectrum component of atom $i$ and $\beta_k$ is the associated linear coefficient, a free parameter in the SNAP model.  
As a computational convenience, the contribution of each bispectrum component to the SNAP energy is shifted by the contribution of an isolated atom, $\beta_k {B}_{k0}^{i}$, so that the SNAP energy of the isolated atom is equal to $\beta_0$ by construction.
Similarly, the force on each atom $j$ due to the SNAP potential can be expressed as a weighted sum over the derivatives w.r.t. ${\bf r}_j$ of the bispectrum components of each atom $i$.
\begin{equation}
{\bf F}^{j}_{\scriptscriptstyle{SNAP}}=-\nabla_{j}\sum_{i=1}^{N}E_{\scriptscriptstyle{SNAP}}^i=-\boldsymbol\beta{\cdot}\sum_{i=1}^{N} \frac {\partial {\bf {B}}^{i}}{\partial {\bf r}_{j}}
\label{eq3}
\end{equation}
In this way, the total energy, forces, and also the stress tensor, can be written as linear functions of quantities related to the bispectrum components of the atoms. 
In addition to shifting the bispectrum components by ${B}_{k0}^{i}$, it also makes sense to set ${\beta}_{0}=0$, constraining the potential energy of an isolated atom to be zero.  
This ensures that SNAP correctly reproduces the cohesive energy of the reference solid structure, an important physical attribute of any general purpose interatomic potential.
For multi-element systems, such as the tungsten-beryllium materials considered here, SNAP captures the effect of compositional differences in several ways.  Firstly,
the coefficients ${\boldsymbol\beta}$ are different for each element.  Secondly, the contributions to the basis functions in Eq. \ref{eq:u} made by each atom depend on the element weight $w_{i'}$ and the effective atomic radius $R_{i'}$.

\subsection{Computational Efficiency}
\begin{figure}[t]
\includegraphics{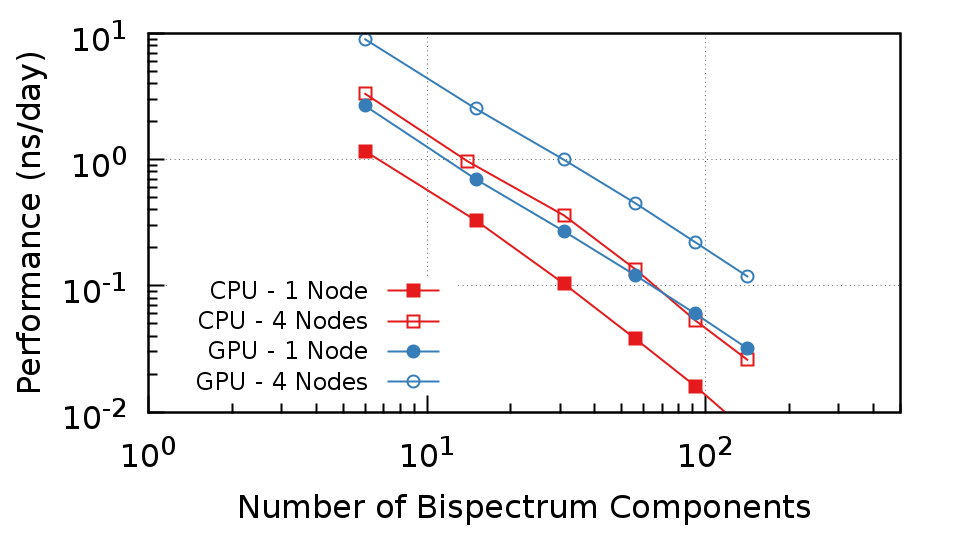}
\caption{\label{timing} Simulation rate (ns/day) for an NVE MD simulation consisting of 31k atoms versus the number of descriptors used in the SNAP potential. The benchmark was run on one and four CPU or GPU nodes. Each CPU node consists of two Intel Broadwell E5-2695 v4 processors, with a total of 36 physical cores per node.  The GPU node consisted of four NVIDIA P100 cards. Both the CPU and GPU systems use an Omni-path interconnect.}
\end{figure}Implementation of SNAP in LAMMPS\cite{plimpton1995fast,lammpsweb} uses the KOKKOS library\cite{trott2014snap,mattox2018highly}, allowing the code to run efficiently on diverse hardware architectures, including CPU, GPU, and many-core processors.  
The implementation also exploits LAMMPS highly-scalable MPI-based spatial decomposition scheme, allowing a single MD simulation to be distributed over a few nodes or an entire supercomputer \cite{trott2014snap}.
The spatial resolution of the potential can be continuously improved by increasing the number of bispectrum components, systematically increasing the accuracy of the SNAP potential, at the price of greater computational cost. 
This is illustrated in Figure~\ref{timing} where the number of bispectrum components is increased from 6 to 141, increasing the computational cost by over two orders of magnitude.
Performance is reported as the amount of MD simulation time that can be calculated in a given amount of wall-clock time (ns/day). 
The data displayed here is for a benchmark problem consisting of 31,250 tungsten atoms, running NVE molecular dynamics with a timestep of 0.5~fs.  
Figure~\ref{timing} compares a traditional CPU (Intel Broadwell) compute node to a modern multi-GPU (four NVIDIA P100's) compute node. 
SNAP scales comparably on either hardware, but there are significant performance gains when multiple GPU cards are assembled onto a single compute node.

\section{\label{train}Training a Machine Learned Model}
\subsection{Constructing the Training Set}
The present work is focused on generating a SNAP interatomic potential for tungsten-beryllium with an intended use in simulating plasma facing components in a fusion reactor.
As such, a training set must be constructed that reflects the material properties relevant to this application space, but also is of general use to end users.
Given the highly flexible nature of machine learned potentials, any reference model can be taken as a training set.
However, we employ SNAP as a multiscale link between density functional theory (DFT) and MD, and as such will need a data base of expensive electronic structure calculations to properly train the model.
Constructing a training set is a critical part of any machine learning endeavor because the constructed model will, by default, be best at interpolating between data it has already seen. 
Therefore, when it comes to an IAP, the more diverse the atomic configurations included in the training set the better suited for \emph{general use} the resultant potential should be.   
Domain size limits within DFT imposes some restrictions of what types of training configurations can be included with an upper limit around a few hundred atoms.
Atomic configurations within these size limitations need to be chosen such that they represent the material properties and application space of interest.
There are no well-defined rules for how best to construct training set for ML-IAP generation.
Physical insight and expert domain knowledge of the materials science application are needed to guide the selection of the DFT atomic configurations.
Alternative methods such as learning \emph{on-the-fly}\cite{csanyi2004learn,li2015molecular,podryabinkin2017active} have been proposed as unsupervised approaches to training set construction, but this is still an area of active research. 

Presently, we have chosen to curate the training set by hand. 
The constructed training set can be divided into three general categories: DFT calculations of pure tungsten, pure beryllium, and those containing both elements.
Table \ref{table:alltrain} lists all of the training data used, as well as the number of energy ($N_E$) and force ($N_F$) points that each group contributes to the overall fit. 
Beginning with the pure tungsten training data, a number of configurations were taken from a data set previously used to fit a GAP potential for tungsten\cite{szlachta2014accuracy, libatoms}. These are the Dislocations, \emph{ab initio} MD, Elastic Deformations, Surfaces, Monovacancies, and two 
$\Gamma$-surface groups. 
Additional DFT calculations were carried out to add the Self-Interstitials, Liquids, Divacancy and Equation of State training groups to the set. 
Pure tungsten training calculations were performed with VASP\cite{kresseInitioMolecularDynamics1993, kresseEfficientIterativeSchemes1996, kresseEfficiencyAbinitioTotal1996} using a 600eV plane wave cutoff energy, 
approx. 0.015 \AA$^{-1}$ (depends on configuration) k-point spacing, a PBE-GGA exchange-correlation 
functional\cite{perdewGeneralizedGradientApproximation1996,blochl1994projector,kresse1999ultrasoft} and a pseudopotential that leaves the outermost s-,p- and d-orbitals to the be solved by the basis set.

While there are any number of additional configurations that could be added, we believe the current training set for tungsten covers most of the bulk behavior (elastic deformations, equation of state, vacancies) as well as high energy configurations that would result from radiation damage(dislocations, interstitials, surfaces).   
In total, there are $9897$ individual atomic configurations in the pure tungsten set, with over $10^{6}$ force data points. 

The beryllium training set has a very similar composition to that of tungsten, since the goal is to create a \emph{general use} potential that is also tailored to simulate plasma facing materials.
Equilibrium bulk properties of beryllium are captured through the Elastic Deformation, Equation of State and \emph{ab initio} MD training groups. Together these groups contribute approximately 95\% and 47\% of the total energy and force training points, respectively. 
Conversely, the defect properties and lower symmetry environments of beryllium are collected in the Surfaces, Self-interstitials, Stacking Fault and Liquid groups. 
While fewer in number of configurations, these large atom count training structures contribute the majority of the force data points.
All of the beryllium training data was also generated using VASP with the same simulation parameters as the tungsten data, with the chosen pseudopotential leaving just the outermost s-orbital electrons to the basis set.  

Lastly, a set of training data was generated that focused on ordered inter-metallic phases of W-Be ranging in composition from equiatomic to WBe$_{12}$.
In addition, multiple crystal structures of these proposed inter-metallic compounds were used in these calculations.
For all training groups except Surface Adhesion, six different phases of W-Be were considered: B$_{2}$ (WBe), L$_{12}$ (WBe$_{3}$), C$_{14}$ (WBe$_{2}$), C$_{15}$ (WBe$_2$), C$_{36}$ (WBe$_{3}$), and D$_{2}$B (WBe$_{12}$).
Surface Adhesion is a special training group that is strongly aligned with the target application of high energy Be implantation onto a W surface. 
This set of configurations included the binding of a single Be atom adsorbed onto $(100)$ and $(111)$ tungsten surfaces as well as multiple Be atoms adsorbed onto the same surface orientations.

The remaining columns in Table \ref{table:alltrain}, $\sigma_{E}$ and $\sigma_{F}$, are the optimal training weights selected for the energies and forces in each training group.
These group weights are scaled by the number of data points in the group, so they indicate the relative importance assigned to
each group in the optimization process.
The bolded values indicate the largest weight in each column, which can be interpreted as the \emph{most important} type of training data for fitting the full W-Be SNAP potential. 
The details of this optimization process and how these optimal training weights were obtained will be discussed in the following section.

\begin{table*}[]
\centering
\resizebox{\textwidth}{!}{%
\begin{tabular}{lcclllcclllccll}
Description&$N_{E}$&$N_{F}$&$\sigma_{E}$&$\sigma_{F}$&Description&$N_{E}$&$N_{F}$&$\sigma_{E}$&$\sigma_{F}$&Description&$N_{E}$&$N_{F}$&$\sigma_{E}$&$\sigma_{F}$\\
\hline
\hline
W:&&&&&Be:&&&&&W-Be:\\
Elastic Deform & 2000 & 6000 & $5\cdot10^{0}$ & $6\cdot10^{4}$ & Elastic Deform & 4594 & 43260 & $1\cdot10^{5}$ & $\bf{1\cdot10^{7}}$ & Elastic Deform$^{\dagger}$ & 3946 & 68040 & $\bf{3\cdot10^{5}}$ & $2\cdot10^{3}$\\
Equation of State & 125 & 3468 & $1\cdot10^{-1}$ & $6\cdot10^{4}$ & Equation of State & 502 & 5418 & $6\cdot10^{4}$ & $3\cdot10^{6}$ & Equation of State$^{\dagger}$ & 1113 & 39627 & $2\cdot10^{5}$ &$4\cdot10^{4}$ \\ 
DFT-MD & 60 & 23040 & $3\cdot10^{0}$ & $1\cdot10^{4}$ & DFT-MD & 909 & 130896 & $\bf{2\cdot10^{5}}$ & $2\cdot10^{6}$ & DFT-MD$^{\dagger}$ & 3360 & 497124 & $7\cdot10^{4}$ & $6\cdot10^{2}$ \\
Surfaces & 180 & 334818 & $\bf{1\cdot10^{5}}$ & $3\cdot10^{5}$ & Surfaces & 90 & 17280 & $1\cdot10^{3}$ & $4\cdot10^{5}$& Surface Adhesion & 381 & 112527 & $2\cdot10^{4}$ & $\bf{9\cdot10^{4}}$ \\ 
Self-Interstitials &15 &5805 & $5\cdot10^{-2}$ & $8\cdot10^{2}$ & Self-Interstitials & 179 & 137931 & $3\cdot10^{2}$ & $4\cdot10^{5}$  & \multicolumn{5}{l}{$\dagger$ multiple crystal phases included in this group:}\\
Liquids & 27 & 3120	&$4\cdot10^{-3}$ & $3\cdot10^{2}$ & Liquids & 75 & 57600 & $7\cdot10^{1}$ & $7\cdot10^{5}$ \\
Dislocations & 98 & 39690 & $3\cdot10^{0}$ & $9\cdot10^{4}$ & Stacking Faults & 6 & 864 & $3\cdot10^{0}$ & $2\cdot10^{6}$ \\
Monovacancy & 420 	& 183054 & $2\cdot10^{3}$ & $1\cdot10^{5}$ &&&&&& \multirow{3}{*}[0.3875in]{ \begin{minipage}{1mm} \includegraphics[]{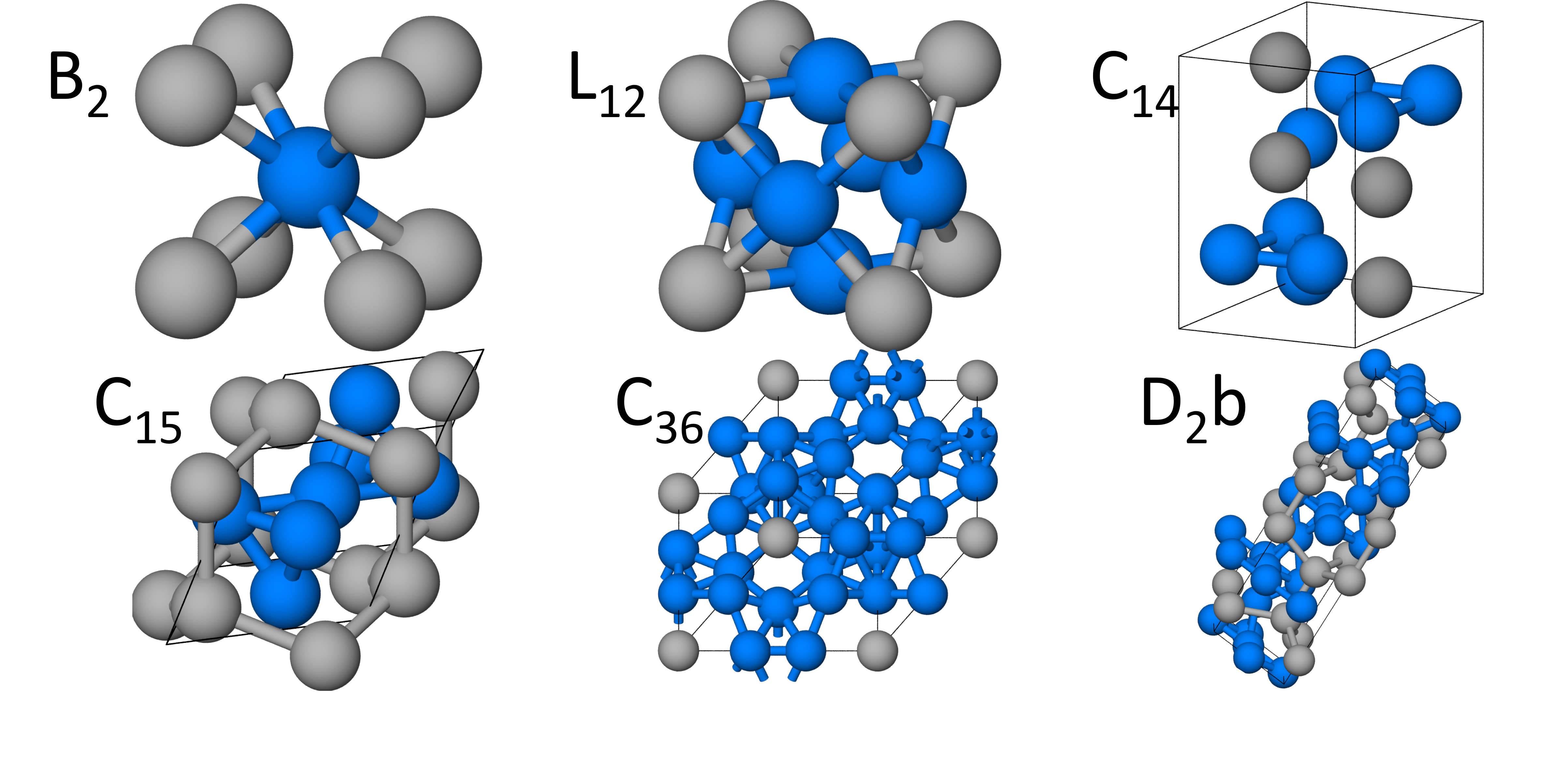} \end{minipage}}\\
Divacancy	 & 39	& 6084 & $1\cdot10^{0}$ & $1\cdot10^{3}$ \\
$\Gamma$-Surface & 6183 & 328338 & $1\cdot10^{0}$ & $1\cdot10^{6}$  \\
$\Gamma$-Surf.+Vacancy & 750 & 105750 & $4\cdot10^{-1}$ & $\bf{3\cdot10^{6}}$ \\
\hline
Total&9897&1039167&&&&6355&393249&&&&8800&717318&&\\
\end{tabular}}
\caption{Training data used in the full W-Be SNAP fit, broken down by element type and group within each element. 
Each of the groups in the W-Be category contain configurations for multiple inter-metallic compounds, some of which are displayed in the inset. For each group the number of energy ($N_E$) and force ($N_F$) training points are given as well as the optimal training weight ($\sigma_E, \sigma_F$) selected for the full W-Be SNAP potential.}
\label{table:alltrain}
\end{table*}

\subsection{\label{optimize}Optimization Methodology}
Once a training set has been constructed, the goal of fitting a SNAP potential is to strike a balance between accurate reproduction of the training data (interpolated properties) and ability to describe structures that are too large to calculate using DFT (extrapolated properties).
The simplest, and most common, interpolation error that can be optimized is the regression error.  
Equation \ref{euclid} captures the general form of linear regression used here.
$\hat{\boldsymbol\beta}$ minimizes the difference between the descriptor ($D$, bispectrum representation) prediction and reference ($T$, electronic structure) data. 
A regularization penalty of order $n$ with weight $\gamma_{n}$ can be applied to constrain the $\hat{\boldsymbol\beta}$ solution.
Solutions with $n=1$ enforce sparsity in the $\hat{\boldsymbol\beta}$ solution, while Tikhonov regularization\cite{tikhonov2013numerical} with $n=2$ penalize against large values of $\hat{\boldsymbol\beta}$ which are hallmarks of an overfit solution.
We have observed no improvement in overall accuracy when enforcing sparsity, and there is little risk of overfitting, because the number of bispectrum descriptors is far less than the number of training points($\mathcal{O}[ 10^{6}]$).
\begin{equation}
\hat{\boldsymbol\beta} = \underset{\boldsymbol\beta}{\operatorname{argmin}} (\|\boldsymbol\epsilon \circ (D \boldsymbol\beta- T) \|^{2}-\gamma_{n}~\|\boldsymbol\beta\|^{n})
\label{euclid}
\end{equation}
Therefore, we solve Equation \ref{euclid} with no regularization penalty, corresponding to the weighted linear least squares solution.

When fitting a SNAP potential, there are two different categories of fitting variables that are controlled by the optimizer.
The first are called hyper-parameters and will directly modify the bispectrum components themselves.
Examples of these are the radial cutoff ($R_{ii^{'}}$), element densities ($w_{i^{'}}$), and cutoff scale factor ($\alpha$) of equations \ref{density} and \ref{mixing}, respectively. 
A second set of fitting variables are the aforementioned group weights, $\boldsymbol\epsilon$, that scale each component of thtarget space ($T$) in equation \ref{euclid}. 
There are far fewer hyper-parameters than group weights with the latter being as numerous as the user sees necessary to divide up the full training set into unique groups.
In order to limit the number of free variables, we have chosen to optimize the hyperparameters and group weights for each element separately before tackling the mixed element training data.

\begin{figure}[t]
\includegraphics{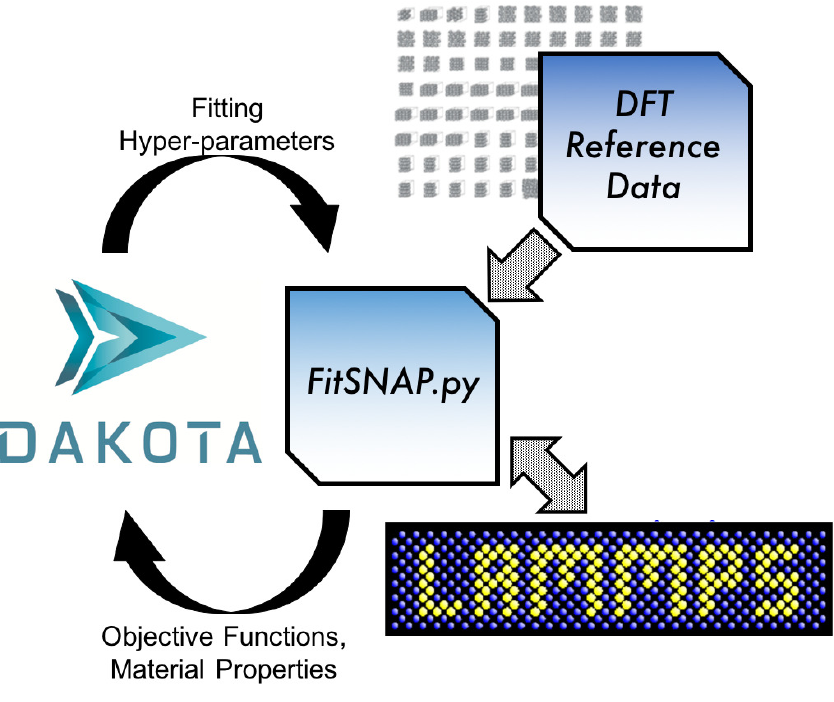}
\caption{\label{flow}Workflow of fitting SNAP potentials. Each of the software tools, DAKOTA, LAMMPS and FitSNAP.py are developed at Sandia National Labs. The reference data can be generated with any code, we have used VASP in the current work.}
\end{figure}

DAKOTA\cite{adams2009dakota} is used as the optimizer utilizing a single objective genetic algorithm (GA). 
Figure \ref{flow} visually displays the overall fitting procedure.
Central to the overall fitting process is FitSNAP.py, which couples DAKOTA, LAMMPS, and the database of DFT training data. 
Following one pass through this optimization loop, a set of fitting parameters is provided from DAKOTA to FitSNAP, new bispectrum components are calculated by taking the coordinate information from the reference data and sending it to LAMMPS.
Once all training configurations are converted into their respective bispectrum components, which forms $D$ of Eq. \ref{euclid}, the energy and forces are parsed from the reference data to populate $T$. 
Solving for $\hat{\boldsymbol\beta}$, the linear regression is done using singular value decomposition and the energy and force errors (interpolation error) are reported back to DAKOTA as part of the objective function.
At this point the candidate potential is used to run short MD simulations to evaluate material properties of interest.
For example, while fitting the tungsten data the elastic constants and a few defect formation energies are calculated for each candidate and their percent error with respect to DFT is communicated back to DAKOTA. 
An equally weighted contribution from each of these material properties plus the regression errors is used to form the objective function for GA optimization.
Each generation of the GA consists of three-hundred candidate potentials.
We observed minimal improvement of the overall fitness after seventy generations.
One advantage of using a GA is that evaluations of candidate potentials (single pass through Figure \ref{flow}) can be done simultaneously.
As a result, the throughput of the overall fitting process can be distributed across a large super computer. 
Asynchronous evaluation tiling is used to distribute each evaluation to separate compute nodes.  
Typical jobs used fifty Intel Broadwell nodes simultaneously to generate and evaluate candidate potentials.


\subsection{\label{interpolation}Interpolated and Fitted Properties}
\begin{figure}[t]
\includegraphics{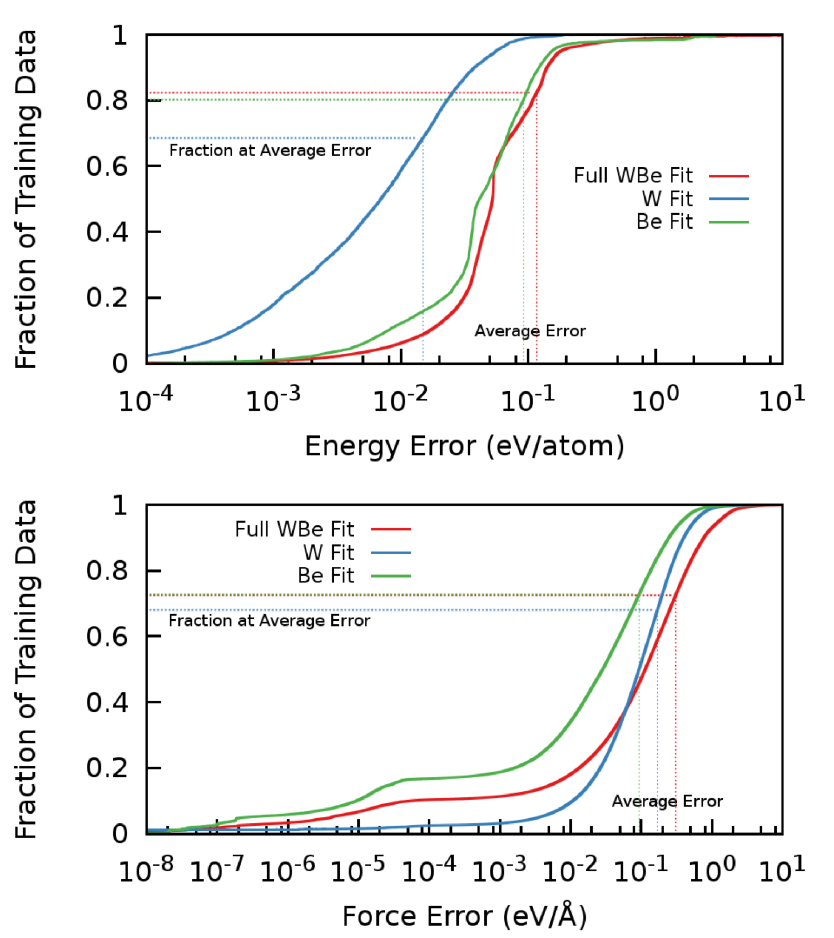}
\caption{\label{InterErrs}Distribution of the regression errors observed for the best fit candidates for each of the pure-W (blue), pure-Be (green) and full W-Be (red) training sets. {\bf (Top)} Energy errors  {\bf (Bottom)} Force errors.  In all cases, vertical dashed lines indicate the average regression error and horizontal dashed lines indicate the fraction of the training data with error lower than the average.}
\end{figure}

Regression errors of the resultant best fit candidate are displayed in Figure \ref{InterErrs}.
The three data series here represent each of the pure phase fits (W Fit and Be Fit, respectively) followed by the optimal fit to the entire W+Be training set.
Due to the increased training set volume and the choice to use linear regression for SNAP potentials, the full W-Be fit has higher average errors, indicated by the dashed vertical lines, than either of the pure component fits. 
However, the fraction of the training data with error below the average is well above 50\%, 
which indicates that the average errors are dominated by a relatively small number of outlier configurations that have exceptionally large energy or force errors.
The average interpolation errors for the fit to the entire W-Be training set are 0.12 eV/atom and 0.31 eV/\AA, respectively.

\begin{figure*}[t]
\begin{centering}
\includegraphics{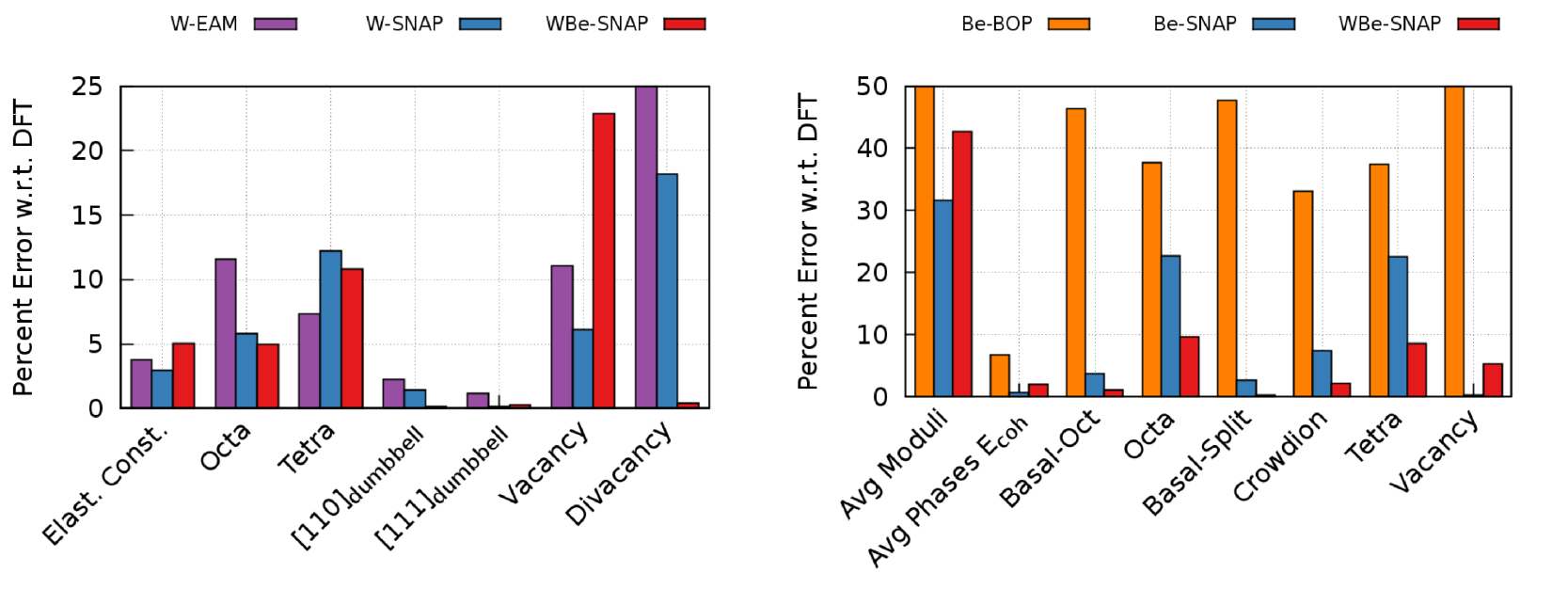}
\caption{\label{DefectErrs}{\bf (Left)} Tungsten material property predictions included in the SNAP optimization loop represented as percent error to the DFT prediction. Due to the small absolute value of the divacancy binding energy (0.12eV) the percent error for EAM is much larger than either SNAP potential displayed here. {\bf (Right)} Beryllium material property predictions that are also included in the optimization procedure. Both SNAP potentials displayed here are significant improvements over the existing BOP predictions of self-interstitial formation energies. }
\end{centering}
\end{figure*}

In addition to these interpolation errors, each candidate potential is used in a set of short MD simulations to determine its accuracy for material properties of interest.
The reference values for these properties are taken from DFT.
The relative errors of these predictions are included in the objective function for hyperparameter optimization.
For tungsten these properties are elastic constants, lattice parameter, cohesive energy, 
and the relaxed formation energies of six point defects in the BCC phase.
Similarly for beryllium the HCP elastic constants, six point defect formation energies, 
and cohesive energies of five simple crystal structures are used as fitting objectives.
These fitted material properties are displayed in Figure \ref{DefectErrs}.
The left and right panels show the percent errors with respect to the DFT predictions for the pure-W and pure-Be SNAP potentials, respectively.  
In both cases, the corresponding results for an empirical potential is shown (EAM\cite{juslin2013interatomic,cusentino2015comparison} for W and BOP\cite{bjorkas2010w} for Be).
One of the primary flaws of the W-EAM potential\cite{juslin2013interatomic} was the prediction of an attractive divacancy binding energy at the nearest neighbor position, whereas DFT predicts\cite{derlet2007multiscale} a mild (0.12 eV) repulsive energy for this defect configuration.
It is believed that vacancy clustering is a key step in surface morphology changes when tungsten is used as a plasma facing component\cite{lhuillier2011trapping}.
Therefore, the sign of the divacancy binding energy plays a critical role in surface evolution.
All of the targeted material properties are within 10\% of the DFT predictions, with the exception of the vacancy formation energy which has an error of 23\% or -0.74eV w.r.t. DFT.

In regard to the beryllium properties, the current SNAP potential is a significant improvement on the existing bond order potential\cite{bjorkas2010w}.
All of the cohesive energies and point defect properties for the present SNAP potential are again within 10\% of DFT.
The exceptions to these positive SNAP results are the HCP elastic moduli.
Our W-Be SNAP potential predicts C$_{13}$ to be -22~GPa whereas DFT predicts\cite{silversmith1970pressure} a value of 17~GPa.
This subsequently makes for a poor description of the shear moduli which captures the basal expansion under compression along the $c$~axis.  

No additional fitting objectives were added for the binary system.
Point defects of dissimilar element species were left as measures of extrapolation accuracy that will be discussed in the following section.

\section{\label{extrapolation}Beryllium Implantation Results}
\begin{figure}[b]
\includegraphics{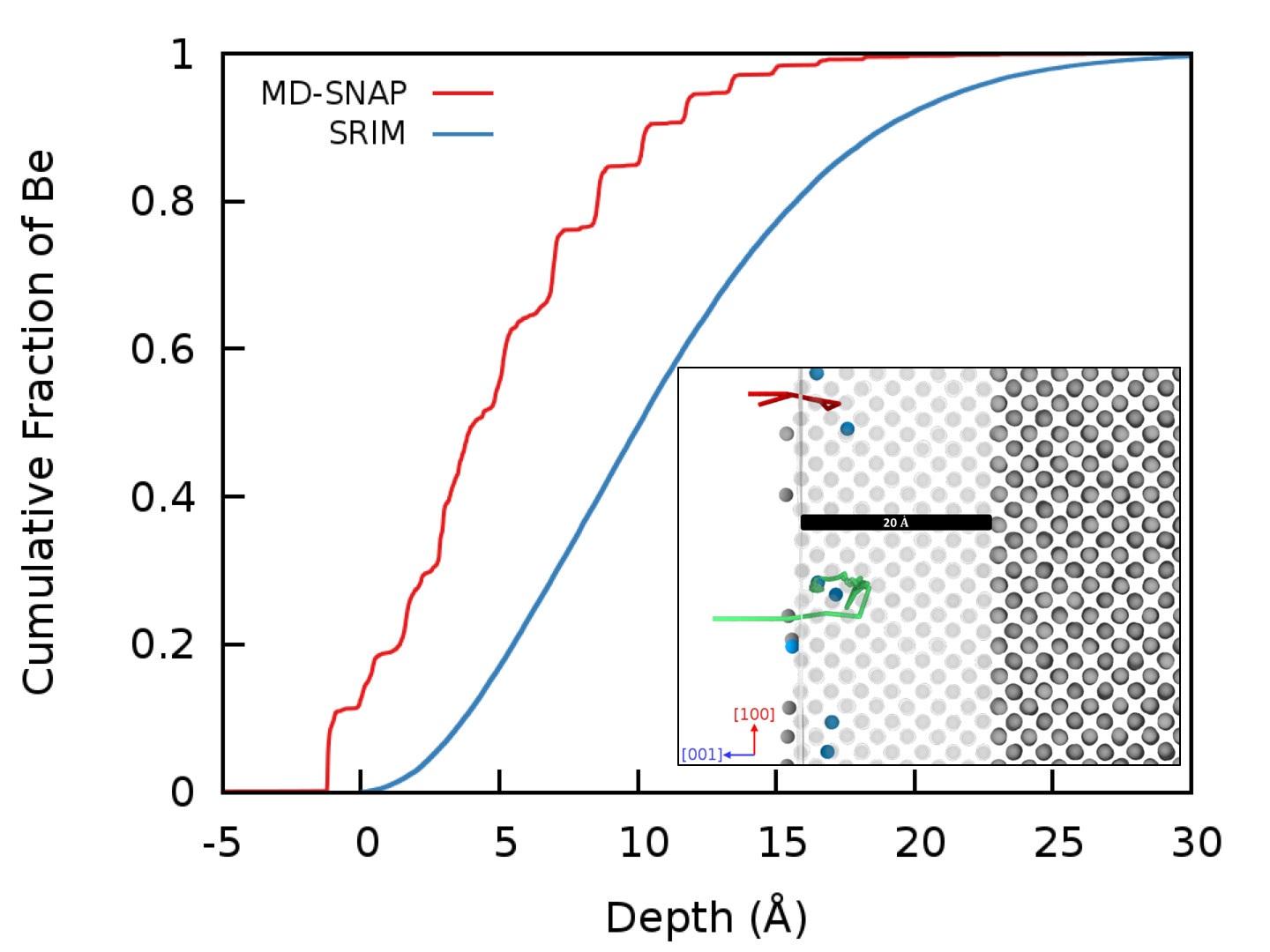}
\caption{\label{depthprof} Plot of the cumulative depth distribution of 75 eV Be in tungsten at 1000 K using both MD (red) and SRIM (blue). Inset displays an atomistic snapshot of the Be (blue spheres) implanted onto the (001) surface of W (grey spheres). The red/green trajectory lines show the time history of a rejected/captured Be atom, respectively. 
}
\label{fig:DepthDist}
\end{figure}
To test the quality of the potential outside of the data included in the training set, molecular dynamics simulations of single beryllium implantations in tungsten were performed. 
Quantifying the implantation depth and lattice interaction of beryllium in tungsten will determine future diffusion and damage mechanisms that will affect overall tungsten diverter performance.  

Simulations were performed using the LAMMPS\cite{plimpton1995fast} molecular dynamics package and the SNAP potential described in this work.  
The simulation cell consisted of a 3 nm x 3 nm x 9 nm tungsten slab with 3 nm of void space above the surface.  
Periodic boundary conditions were used in the $x$, [100] and $y$, [010] directions while a free surface boundary condition was used in the $z$, [001] direction.  
The tungsten was first equilibrated to a temperature of 1000 K by giving the atoms a velocity based on the Maxwell Boltzmann distribution.  
Dynamics were run with an NVE thermostat and a 1 fs timestep for 20 ps where velocity rescaling was performed for the first 5 ps and then turned off for the last 15 ps.  
After equilibration, a beryllium atom was placed {10~\AA} above the surface with random $x$ and $y$ coordinates.  
The beryllium atom was then given an energy of 75 eV in the $z$ direction directly towards the surface and dynamics were performed with an NVE thermostat.  
During the implantation, a variable timestep was required to conserve energy due to the initially high beryllium velocity.  
The timestep was allowed to vary between $10^{-4}$ fs and 0.5 fs and was updated every 10 timesteps so that no atom moved more than {0.02~\AA} per time step.  
It was necessary to freeze the bottom two layers of atoms by setting their forces to zero to prevent the unwanted movement of the slab.  
The simulation was allowed to evolve for 3 ps and the beryllium location in the lattice was subsequently recorded unless it reflected from the surface.   
A total of 5,000 individual simulations were performed.  
A similar calculation for 75 eV beryllium implantation in tungsten was run in SRIM\cite{ziegler2010srim} for $1\cdot10^{6}$ atoms for comparison.


Of the total beryllium implantations performed with the newly generated SNAP potential, 35\% implanted in the lattice while the other 65\% reflected.  
A plot of the beryllium depth distributions for SNAP and SRIM\cite{ziegler2008srim} are shown in Figure \ref{fig:DepthDist} in red and blue respectively.  
The SNAP potential predicts the beryllium atoms to remain within {20~\AA} of the surface after implantation with about 12\% of the beryllium atoms residing above the original surface, indicating a preference for the beryllium to be near the surface.    
While the SRIM profile is comparable to SNAP, SRIM predicts a slightly deeper depth profile and a lower reflection rate.
Nevertheless, both distributions indicate that implanted beryllium remains near the surface after implantation.  
The distinct stepped profile produced by SNAP reflects the tendency of the beryllium atoms occupy
particular interstitial sites within the tungsten matrix.
SRIM does not capture this effect, since the material is modeled as a homogeneous isotropic material with no crystalline structure.  
While SRIM also includes electronic stopping and MD does not, the depth profiles is still more shallow for the SNAP potential.
This indicates the importance of atomic collisions in the beryllium implantation process.  
Overall the two distributions are fairly consistent and differ only slightly.
This is most likely due to the different assumptions used in the two different methods.

The inset in Figure \ref{fig:DepthDist} depicts the beryllium trajectory for a few different individual implantations.  
The red line traces the history of a beryllium atom that entered the lattice but subsequently escaped while the green line traces a beryllium atom that implanted.  
Captured beryllium diffuses rapidly during the brief thermalization process, as indicated by the jagged trajectory line,
and eventually becomes trapped just under the surface.  
The impacting beryllium atoms interact with the tungsten lattice in a variety of ways including displacing a tungsten atom and subsequently occupying the vacant site, creating tungsten adatoms (see inset image of Figure \ref{fig:DepthDist}), creating W-Be dumbbells, and sputtering tungsten atoms with a low sputtering yield of 0.006 W/Be.    

Initial observations of the simulations indicated that implanted beryllium atoms typically resided in interstitial sites, substitutional sites, surface sites, or as $\left<111\right>$ or $\left<110\right>$ oriented W-Be dumbbells.  
For the case of $\left<111\right>$ W-Be dumbbell formation, the configuration is more like a series of oriented displacements in the $\left<111\right>$ direction with a beryllium at the center and typically two displaced tungsten atoms.  
Beryllium that substitutes a tungsten atom on the lattice results in the displaced tungsten atom typically residing on the surface as an adatom.  
All of the 12\% of the beryllium atoms above the surface in the depth profile were identified to be at hollow sites.   
The number of implanted beryllium atoms at each site was quantified by extracting the lattice position and is listed in Table \ref{table:FormationEnergy}.   
Overall the beryllium atoms preferred the $\left<111\right>$ dumbbell, followed by the substitutional site and the hollow site on the surface.

\begin{table}[t]
\begin{tabular}{p{2.5cm} p{1.5cm} p{0.25cm} p{1cm} p{1cm} p{1cm}}
&Implanted&&\multicolumn{3}{c}{Formation Energy (eV)}\\
Defect Type&Be Percent&&DFT&SNAP&BOP\cite{bjorkas2010w}\\
\hline
\hline
$[111]$ Dumbbell &~~~~41.2&&~4.30&~3.66&~0.67\\
Substitution &~~~~22.2&&~3.11&~3.29&-2.00\\
Surf. Hollow Site &~~~~12.3&&-1.05&-1.39&-3.52\\
Tetrahedral Inter.&~~~~10.4&&~4.13&~4.20&-0.28\\
$[110]$ Dumbbell &~~~~8.4&&~4.86&~4.29&-0.03\\
Octahedral Inter.&~~~~5.3&&~3.00&~5.11&~0.34\\
Surf. Bridge Site &~~~~0.03&&~1.01&~0.44&-1.30\\
\hline
\end{tabular}
\caption{Defect formation statistics for single, 75~eV, Be implantation onto a (001) surface of BCC tungsten with a comparison of formation energies for these Be interstitials in the W matrix. While SNAP was only trained for self-interstitial energies for either element type, its prediction of these multi-element defects are much closer to DFT than the empirical BOP potential.\cite{bjorkas2010w}}
\label{table:FormationEnergy}
\end{table}

To determine how realistic the rate of occurrence of these beryllium interstitials are, a series of new DFT calculations has been performed to assess these defect formation energies.  
It is important to note that these formation energies were not included in the training data and are therefore a good test of how well
the potential can predict properties relevant for this particular application. 
Values of the defect formation energies calculated using DFT and SNAP, as well as the existing BOP potential for comparison, 
are shown in Table \ref{table:FormationEnergy}.  
The SNAP potential performs very well for most cases, wth the exception of the octahedral formation energy and the $\left<111\right>$ dumbbell.  
Nevertheless, the new SNAP potential is more consistent with DFT values than is BOP.  
Furthermore, SNAP predicts the three lowest formation energies to be the surface hollow site, the substitutional site, and the $\left<111\right>$ dumbbell.  
These three defects are also the most frequently observed defects in the implantation simulations, indicating consistency between the MD results
and the defect formation energy calculations. 
While the surface hollow site has the lowest formation energy, the $\left<111\right>$ dumbbell as well as the substitutional defect are observed more frequently.  
This is due to the beryllium being implanted with a kinetic energy of 75~eV, where it is more likely that beryllium will sample the sub-surface defect types than to diffuse back up to the surface during the implantation process.   
While there is some discrepancy between SNAP and DFT for these defect formation energies, the agreement is quite good given that the training data was focused on ordered intermetallic phases of W-Be and not low-symmetry atomic configurations resembling defects.

\section{\label{discuss}Discussion}
These initial MD simulation results provide a first evaluation of the implanted beryllium profile, as well as identifying the initial fate of the beryllium once in the lattice. 
This advanced ML-IAP enables larger MD simulations that can be used to investigate longer time scale ($\mathcal{O}[ 10^{-1}-10^{1}]\mu s$) evolution of the tungsten surface subjected
to beryllium implantation. 
These simulations will reveal important physics related to the timescale associated with W-Be intermetallic phase formation, as well as local defect configurations that may serve as trapping sites for implanted hydrogen or helium atoms. 
Large-scale MD simulations can also provide important computational data for benchmarking longer time mesoscale or continuum simulation techniques.

The plasma-surface interactions (PSIs) occurring in the diverter and plasma facing components (PFCs) pose a critical scientific challenge that limits our ability to operate fusion machines by sustaining a steady-state burning plasma. 
The simulation paradigm of multiscale computational modeling relies on a parameter-passing framework in which the entire spatial and temporal domains are sub-partitioned into different regimes on the basis of the characteristic length and time scale of the physical phenomena involved. 
Such multiscale models attack the complex materials degradation issues from both a \emph{bottom-up} atomistic-based approach simultaneously with a \emph{top-down} continuum perspective, and focus on the hierarchical integration of kinetic processes of species reactions and diffusion to model microstructure evolution over experimental timescales. 
The simultaneous use of both an atomistic and continuum approach has furthered the development of scale-bridging or multi-scale integration, and has led to fundamental insight into helium-hydrogen synergies controlling PSI in tungsten, as well as the long-term microstructural evolution due to radiation damage in structural materials.\cite{marian2017recent} 

First-principles, density functional theory (DFT) electronic structure methods as implemented in commercial and open-source codes\cite{kresse1993ab,giannozzi2009quantum,soler2002siesta} can be instrumental in providing interaction forces, basic thermodynamic and kinetic interactions and rates, which can be used in fitting interatomic potentials for molecular dynamics simulations, and are utilized where existing interatomic potentials are deemed inadequate.
Unfortunately the limitation of such first principles methods relate to the lack of thermal fluctuations in DFT calculations of thermodynamics and migration barriers, as well as the very short timescales ($\mathcal{O}[10^{2}]$~ps) available for dynamic DFT-MD simulations. 
Moving past the size and time limitations of DFT, large-scale MD simulations can provide an extension to the \emph{bottom up} multiscale modeling paradigm. 
MD simulations are only as accurate as the interatomic potentials, but can provide important physical insights on the dynamics of defect interactions, provided that such interaction dynamics occur on rapid, nanosecond timescales. 

Furthermore, MD simulations can provide a computational database capable of benchmarking mesoscale or continuum scale models, as well as identifying key physical mechanisms that must be included in longer-time simulation techniques.
 The emerging multiscale modeling capabilities are very much in the early stages of development, and continued research activities are required to further develop this capability. 
 In particular, the questions around mixed material formation including the timescale on which intermetallic phase separation occurs, how such phases and localized chemically complex defect arrangements influence hydrogen retention and permeation, require atomistic insight. 
 These initial MD simulations, and the improvements in modeling chemically complex plasma exposed surfaces using the SNAP interatomic potentials, provide a key opportunity to investigate such complex and important PSI challenges.

\section{\label{conclusions}Conclusions}
At the intersection of data-science and atomistic simulation of materials, the presented ML-IAP demonstrates the significant improvement over empirical IAP that can be provided by SNAP.
This new SNAP W-Be potential improves upon the existing BOP for key material properties that are necessary for studying PFCs and ultimately this accuracy and scalability improvement will become a key component of multiscale simulation of PFCs. 
The results of the Be implantation simulations discussed here indicate a preference for surface adhesion and shallow depth profiles into tungsten.
This SNAP W-Be potential will allow for further simulations targeting W-Be plasma material interactions, filling a critical need in the area of fusion energy research.
The results presented here show consistency with DFT for important defect properties relevant to Be implantation in W.
How these implantation defects affect helium and hydrogen trapping from the plasma, as well as long timescale dynamics of Be at W surfaces is the focus of future work.
Lastly, the fitting methodology outlined here can be safely applied to any condensed phase system given suitable training data, though additional study is needed to validate the bispectrum as a physical descriptor of gaseous and molecular bonding environments.




\begin{acknowledgments}

The new SNAP W-Be potential has been added to the public distribution of the LAMMPS software package \cite{lammpsweb}.  The training data is available upon request from the authors.

All authors gratefully acknowledge funding support is from the plasma surface interaction project of the Scientific Discovery through Advanced Computing (SciDAC) program, which is jointly sponsored by the Fusion Energy Sciences (FES) and the Advanced Scientific Computing Research (ASCR) programs within the U.S. Department of Energy Office of Science.
Equal support of this work is from the Exascale Computing Project (No. 17-SC-20-SC), a collaborative effort of the U.S. Department of Energy Office of Science and the National Nuclear Security Administration.
Sandia National Laboratories is a multi-mission laboratory managed and operated by National Technology and Engineering Solutions of Sandia, LLC, a wholly owned subsidiary of Honeywell International, Inc., for the U.S. Department of Energy's National Nuclear Security Administration under contract DE-NA0003525.
\end{acknowledgments}

\bibliography{arxiv_Submit_Feb25}




\end{document}